\documentclass[prb,aps,superscriptaddress,floatfix,tightenlines,showpacs,twocolumn]{revtex4-2}

\usepackage{graphicx}
\usepackage{dcolumn}   
\usepackage{bm}        
\usepackage{amssymb}   
\usepackage{amsmath}
\usepackage{url}
\usepackage{layout}
\usepackage{color}
\usepackage{epsfig}
\usepackage{graphicx}
\usepackage{booktabs}
\usepackage{float}
\usepackage{hyperref}
\hypersetup{colorlinks,citecolor=blue,linkcolor=blue,filecolor=magenta}

\begin{document}

\title{Spin Quadrupolar orders in $d$-wave Unconventional Magnetism}

\author{Jian-Keng Yuan}
\affiliation{Fudan University, Shanghai 200433, China}
\affiliation{New Cornerstone Science Laboratory, Department of Physics, School of Science, Westlake University, Hangzhou 310024, Zhejiang, China}
\author{Zhiming Pan}
\email{panzhiming@xmu.edu.cn}
\affiliation{Department of Physics, Xiamen University, Xiamen 361005, China}
\author{Congjun Wu}
\email{wucongjun@westlake.edu.cn}
\affiliation{New Cornerstone Science Laboratory, Department of Physics, School of Science, Westlake University, Hangzhou 310024, Zhejiang, China}
\affiliation{Institute for Theoretical Sciences, Westlake University, Hangzhou 310024, Zhejiang, China}
\affiliation{Key Laboratory for Quantum Materials of Zhejiang Province, School of Science, Westlake University, Hangzhou 310024, Zhejiang, China}
\affiliation{Institute of Natural Sciences, Westlake Institute for Advanced Study, Hangzhou 310024, Zhejiang, China}
\date{\today}

\begin{abstract}
Unconventional magnetism  represents a class of metallic states whose Fermi surfaces exhibit spin-dependent splittings under the non-trivial representations of the rotation group. 
The $d$-wave $\alpha$-phase unconventional magnetic state, commonly known as altermagnet,
recently, has attracted significant attention. 
While these systems exhibit distinct anisotropic $d$-wave characteristics in momentum space, how this microscopic topology translates into 
the spin distributions in real space remains a question. 
In this work, we bridge the intrinsic spin quadrupolar ordering in momentum space to the real-space staggered magnetic distribution.
By introducing a weak, non-magnetic periodic crystal potential into a $d$-wave unconventional magnetic state, the spin-charge cross susceptibility is calculated by using the linear response theory. 
We reveal that the interplay between the crystal potential and the intrinsic $d$-wave spin-splitting naturally induces a spatial spin quadrupole distribution without enlarging the unit cell. 
Our study thus provides a physical connection between momentum-space multipoles in the even partial wave channel and real-space spin multipole orders.
\end{abstract}

\date{\today}
\maketitle

\textit{Introduction.} 
Itinerant ferromagnetism exhibits a uniform spin polarization over the Fermi surface, whose symmetry is analogous to conventional $s$-wave superconductivity. 
Similarly, drawing inspiration from higher partial-wave pairing in superconductivity, 
the concept of magnetic ordering can also be extended to the spin channel when spin-orbit coupling (SOC) is negligible ~\cite{wu2004soc,wu2007fermi}.
This phenomenon, termed ``unconventional magnetism'' (UM) ~\cite{wu07website,lee2009}, forms non-trivial representations of the rotational group. 
This phase exhibits an intrinsic, anisotropic distortion of the Fermi surface, leading to the orthogonal elliptical pockets for opposite spin sectors, thereby establishing a spin-splitting effect independent of relativistic SOC.
A prominent example is the collinear $\alpha$-phase in the $d$-wave channel, which breaks time-reversal symmetry but maintains parity ~\cite{wu2007fermi,oganesyan2001quantum,jungwirth2025altermagnetism}.

Recently, unconventional magnetism, especially the $\alpha$-phase in even partial-wave channels, has attracted great attention under the name of ``altermagnetism" ~\cite{smejkal2022emerging,smejkal2022beyond,mazin2022alter}.  
In altermagnets, the real space periodicity of the magnetic structure is the same as the underlying crystalline one.
It also exhibits a spin-group type symmetry, {\it i.e.}, a combination of different types of orbital and spin rotations.
This symmetry classification, reliant on the framework of spin-groups ~\cite{liu2022spin,xiao2024spin,chen2024enumeration,jiang2024enumeration,cheong2025alter}, systematically organizes these magnetic orders and dictates their transport behaviors. 
A common example of spin group symmetry is a spatial rotation around the $z$-axis at the angle of $\frac{\pi}{2}$ followed by a $\pi$-rotation of spin. 
Unlike antiferromagnet, this spin structure results in spin-split Fermi surfaces in the absence of macroscopic net magnetization. 
The interplay between the electron band structure and lattice-compatible magnetic orders has proven to be incredibly rich, leading to novel spintronic applications ~\cite{song2025fun,duan2025antiferro,wang2025soc,zhou2025man,li2026man,tian2025spin,jaeschke2025atomic,bhattarai2025high}, anomalous Hall effects~\cite{feng2022anomalous,sato2024altermagnetic,attias2024intrinsic}, and topological phenomena.

A natural question raises: What is the connection between the $d$-wave unconventional magnets and the altermagnet?
The $d$-wave UM exhibits spin quadruple order in momentum space which does not break translation symmetry as in the usual case of spin-density-wave state.
On the other hand, the altermagnet shows spin quadruple order in the real space without enlarging the unit cell, which does not break lattice translation symmetry. 
Furthermore, their quadruple orders share the same symmetries under rotation, time-reversal and inversion transformations, i.e., they belong to the same symmetry class, 
hence, it is reasonable to expect that they can be smoothly evolve to each other in the presence of lattice potential. 
Nevertheless, how can they be connected  has not been clearly explained. 


In this work, we show that the momentum-space spin quadruplar order generates the real space spin quadruplar order within one unit cell upon the appearance of lattice potential.
A weak periodic crystal potential is introduced into the metallic state whose Fermi surface exhibits the $d$-wave unconventional magnetic order.  
Under the frame work of linear response, the static spin-charge cross-susceptibility is calculated to determine the induced real-space spin distribution analytically.
Scattering from the non-magnetic periodical potential naturally gives rise to a real space staggered spatial spin modulation whose spatial periodicity is the same as the that of the crystal potential. 
More importantly, we establish an equivalence showing that this cross-susceptibility is linearly proportional to the spin quadrupole 
ordering in momentum space. 
Consequently, the $d$-wave $\alpha$-phase of unconventional magnetism is smoothly connected to the altermagnetic states by the adiabatic evolution.


\textit{$d$-wave magnetism in crystals.} 
In 2D, the symmetry-dictated low-energy effective Hamiltonian for a $d$-wave $\alpha$-phase unconventional magnetism could be typically formulated by a continuum model,
\begin{equation}
H_{0}=-\frac{1}{2m} [ \nabla^{2} +Q(\partial_{x}^{2}-\partial_{y}^{2})\sigma_{z}] -\mu,
\label{eq:Hamiltonian0}
\end{equation}
where the Pauli matrix $\sigma_{z}$ acts in the
spin space, $m$ is the effective mass, and $\mu$ is the chemical potential.
Here, we have set $\hbar=1$ for simplicity.

\begin{figure}[t]
\centering
\includegraphics[width=0.45\linewidth]{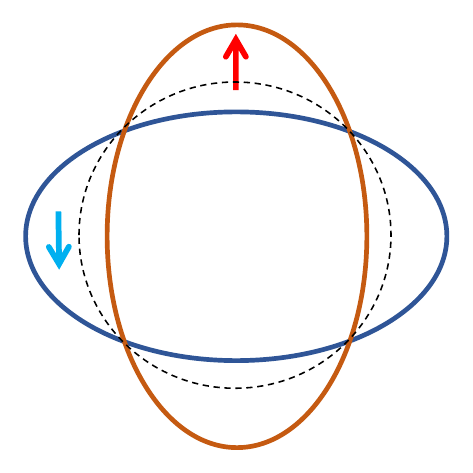}
\caption{The Fermi surface of $d$-wave $\alpha$-phase unconventional magnetic states. The distortion Fermi surface is given by the Hamiltonian in Eq.~(\ref{eq:Hamiltonian0}) and reflects the specific symmetry of $d$-wave magnetic states in Eq.~(\ref{symm_spectrum}).   
}
\label{fig:fermi_surface}
\end{figure}

The first term represents the isotropic kinetic energy, while the second term captures the $d$-wave UM order.
In momentum space, the low-energy effective spectrum of the $d$-wave $\alpha$-phase in Eq.~(\ref{eq:Hamiltonian0}) is given by,
\begin{equation}
\xi_{\sigma}(\bm{k})= \frac{1}{2m} 
[\bm{k}^{2} +Q(k_{x}^{2}-k_{y}^{2})
\sigma -k_{f}^{2}],
\label{spectrum}
\end{equation}
where $\sigma=\pm1$ denotes the spin up and spin down projections, respectively, and $k_{f}$ is the effective Fermi momentum characterizing the unperturbed chemical potential $\mu=k_{f}^{2}/2m$. 
Crucially, this energy spectrum dictates a specific symmetry under a $\pi/2$ rotation around the $z$-axis, as shown in Fig.~\ref{fig:fermi_surface}.
This order breaks both time-reversal ($\mathcal{T}$) and $C_4$ rotational symmetries but preserves the combined $C_4\mathcal{T}$ symmetry, resulting in a spin-dependent anisotropic effective mass.
The dimensionless parameter $Q$ characterizes the strength of the $d$-wave magnetic order.
This is constrained to the range $-1<Q<1$ to ensure that the kinetic energy remains positive-definite along all spatial directions, thereby maintaining the stability of the system at the $\Gamma$ point.

While Eq.~(\ref{eq:Hamiltonian0}) successfully captures the spin-splitting  of altermagnetism in a continuum, it lacks the Brillouin zone topology inherent to real crystals. 
To investigate the interplay between the $d$-wave spin-splitting and lattice translation symmetry, such as band folding and gap openings at the zone boundaries, we adopt a nearly free electron (empty lattice) approach. 
For simplicity, we subject this continuum model to a weak periodic crystal potential of a 2D square lattice, expressed in real space as:
\begin{equation}
V(x,y) = V_{0} (\cos Gx+\cos Gy ),
\label{crystal_potential}
\end{equation}
where $V_0$ is the potential strength and $G$ is the magnitude of the primitive reciprocal lattice vector. 
Through Fourier transformation, the potential in momentum space acts as a scattering center that transfers momentum by reciprocal lattice vectors:
\begin{equation}
V(\bm{q})= \frac{V_{0}}{2} \sum_{\bm{G}\in \{\pm \bm{G}_1,\pm\bm{G}_2 \} } \delta_{\bm{q},\bm{G}} 
\end{equation}
where $\bm{G}_{1}=(G,0)$ and $\bm{G}_{2}=(0,G)$. 
This phenomenological framework allows us to qualitatively capture the generic Fermi surface evolution and topological properties of $d$-wave magnetic states in a crystal without resorting to a complex tight-binding model.

\textit{Spin Response.}
While the spin-density waves could arise from Fermi surface nesting or strong correlations, a spatial modulation of spin density can also emerge in a $d$-wave $\alpha$-phase (altermagnetism). 
Due to the intrinsic distortion of the spin-split Fermi surfaces, a purely non-magnetic scalar potential can directly induce a static spin density modulation.
Here, we employ the linear response theory to calculate this spin-charge cross-response induced by a weak periodic crystal potential.

\begin{figure}[t]
\centering
\includegraphics[width=0.65\linewidth]{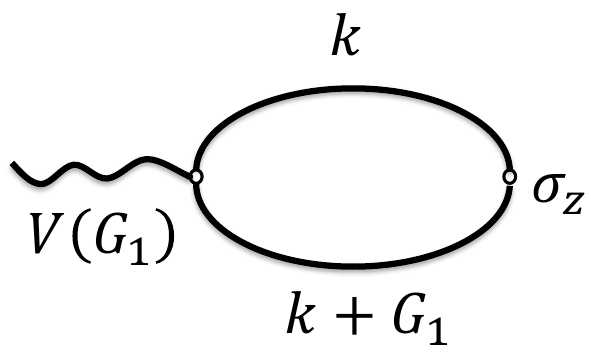}
\caption{The static spin-charge cross susceptibility $\chi(\bm{G_1})$ bubble.  
}
\label{fig:Bubble}
\end{figure}

To determine the induced real space spin distribution, we evaluate the static spin-charge cross-susceptibility. 
The bare Green's function is diagonal in spin space:
\begin{equation}
G(\bm{k},i\omega_{n}) = \frac{1}{2}(
\frac{1+\sigma_{z}}{i\omega_{n}-\xi_{+}(\bm{k})} 
+\frac{1-\sigma_{z}}{i\omega_{n}-\xi_{-}(\bm{k})}),
\end{equation}
where $\omega_{n}$ represents the fermionic Matsubara frequencies.
The correlation function $\chi$ between the $z$-component of the spin density and the charge density at a reciprocal lattice vector $\bm{G}_1$ is as (Fig.~\ref{fig:Bubble}),
\begin{equation}
\begin{aligned}
& \chi(\bm{G}_{1}) \\
= & T\sum_{i\omega_{n}}\int_{\text{BZ}}\frac{d ^{2}\bm{k}}{(2\pi)^{2}}\text{tr}[G(\bm{k},i\omega_{n})G(\bm{k}+\bm{G}_{1},i\omega_{n})\sigma_{z}] \\
= & \int_{\text{BZ}}\frac{d ^{2}\bm{k}}{(2\pi)^{2}}\sum_{\sigma=\pm1}\sigma\frac{n_{F}[\xi_{\sigma}(\bm{k})]-n_{F}[\xi_{\sigma}(\bm{k}+\bm{G}_{1})]}{\xi_{\sigma}(\bm{k})-\xi_{\sigma}(\bm{k}+\bm{G}_{1})},
\end{aligned}
\label{def_chi}
\end{equation}
where BZ denotes the first Brillion zone and $n_{F}(\epsilon)$ is the Fermi-Dirac distribution. 
Defining the rotation operator $R$ such that $R(k_x,k_y)=(-k_y,k_x)$,
the spectrum satisfies:
\begin{equation}
\xi_{\sigma}(R\bm{k})
=\xi_{-\sigma}(\bm{k}).
\label{symm_spectrum}
\end{equation}
Under the rotation transformation, $\bm{k}\rightarrow R\bm{k}$, the Brillion zone of the square lattice remains invariant.
By utilizing the spectral symmetry in Eq.~(\ref{symm_spectrum}) and transforming the integration variables, one can rigorously show that the cross-susceptibility is antisymmetric under the $C_4$ rotation:
\begin{equation}
\chi(R\bm{G}_{1}) =-\chi(\bm{G}_{1}). 
\label{symm_chi}
\end{equation}
Given that $\bm{G}_2= R \bm{G_1}$, and assuming time-reversal/inversion properties ensure $\chi(\bm{G})=\chi(-\bm{G})$, the susceptibilities satisfy the relation:
\begin{equation}
\chi(\pm\bm{G}_{1}) = -\chi(\pm\bm{G}_{2}) =\chi.
\label{realtion_chi}
\end{equation}

\begin{figure}[t]
\centering
\includegraphics[width=0.65\linewidth]{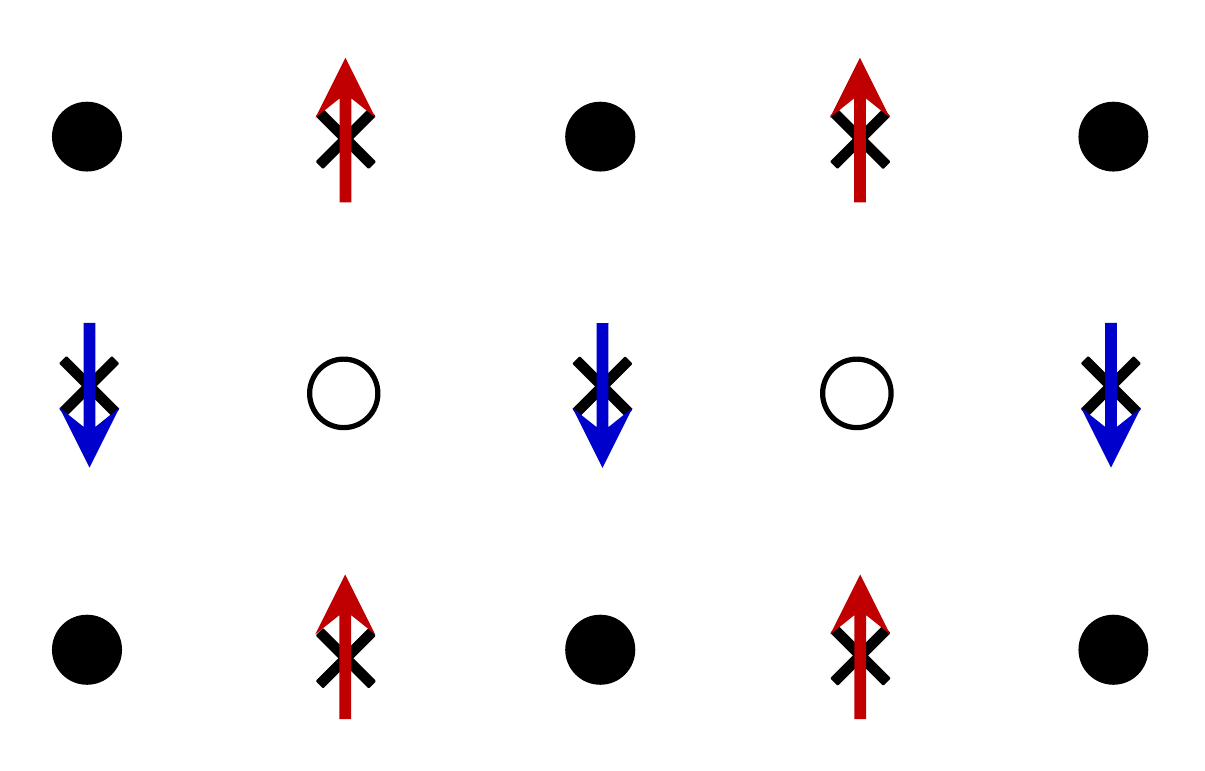}
\caption{The crystal potential (Black) and spin response (arrows) in $d$-wave unconventional magnetic states. The Black solid circles, open circles, and the crosses represent the maximums, minimums, and saddle points of the crystal potential (i.e., where $V(\bm{r})=0$), representatively. The red and blue arrows denotes the maximums and minimums of the spin response. 
}
\label{fig:spin_response}
\end{figure}

Summing over the Fourier components of the crystal potential in Eq.~(\ref{crystal_potential}), the induced real-space spin density up to the first-order correction is derived as:
\begin{equation}
S_{z}(\bm{r}) =  \chi V_{0} (\cos Gx-\cos Gy). 
\label{SDW}
\end{equation}
The induced spin response is a synergistic consequence of the pristine crystal potential and the intrinsic $d$-wave UM order.
Comparing the spatial profile of the periodic potential $V(\bm{r})$ with the induced spin density $S_z(\bm{r})$, we observe that the maximum and minimum magnitudes of the spin response occur at the saddle points of the underlying periodic potential (i.e., where $V(\bm{r})=0$), as shown in Fig.~\ref{fig:spin_response}.

In the paramagnetic normal state (where $Q=0$), the cross-susceptibility $\chi$ vanishes, and consequently, no spin response is induced. 
To quantitatively elucidate the precise role of the $d$-wave UM order in driving this spin response, we evaluate the susceptibility 
$\chi$ as a function of the order parameter $Q$.

To make the discussion concrete, we focus on the low-electron-density regime at zero temperature. 
In this limit, the Fermi momentum is small, and the Fermi surfaces remain tightly confined around the $\Gamma$ point, far from the boundaries of the Brillouin zone. 
The relevant electron momenta satisfy $|\bm{k}|\ll G/2$. 
By expanding the energy denominator up to the leading order in $|\bm{k}|/G$, we obtain:
\begin{equation}
\frac{1}{\xi_{\sigma}(\bm{k}) -\xi_{\sigma}(\bm{k}+\bm{G}_{1})} 
= -\frac{2m}{G^{2}(1+\sigma Q)}.
\end{equation}
At zero temperature, the Fermi-Dirac distribution reduces to a step function, $n_{F}(\epsilon)=\Theta(-\epsilon)$, where $\Theta(\epsilon)=0$ for $\epsilon\leq0$ and $\Theta(\epsilon)=1$ for $\epsilon>0$. 
The susceptibility according to Eq.~(\ref{def_chi}) is 
\begin{equation}
\chi =  -\frac{mk_{f}^{2}}{\pi G^{2}}\frac{1}{\sqrt{1-Q^{2}}}(\frac{1}{1+Q}-\frac{1}{1-Q}).
\label{eq:chi}
\end{equation}

The analytical expression captures the underlying physics. 
As expected, the $d$-wave UM order distorts the initially isotropic Fermi surface into spin-dependent ellipses.
This $d$-wave anisotropy yields a non-vanishing cross-susceptibility, converting the purely scalar crystal potential into the periodic spin density modulation described in Eq.~(\ref{SDW}). 
The susceptibility $\chi$ in Eq.~(\ref{eq:chi}) is proportional to $Q$ for weak ordering and grows monotonically as the order parameter increases. 
The emergence of the $d$-wave magnetic phase ($Q\neq 0$) acts as the driver for generating this spatially modulated spin response.
Moreover, the cross-susceptibility tends to be infinite in the limit of $Q\rightarrow \pm1$. 

\textit{Quadrupole and Susceptibility.-- }
To elucidate the physical origin of the induced spin response, it will be instructive to connect the macroscopic cross-susceptibility to the intrinsic topological and geometric properties of the unperturbed altermagnetic ground state. 
We achieve this by examining the momentum-space spin quadrupole moment, which characterizes the $d$-wave spin-spitting. 
It is defined as:
\begin{equation}
Q_{k_{\mu}k_{\nu}} = \int_{\text{BZ}}\frac{d ^{2}\bm{k}}{(2\pi)^{2}} 
\sum_{\sigma=\pm1} \sigma n_{F}[\xi_{\sigma}(\bm{k})] k_{\mu}k_{\nu},
\end{equation}
where $\mu,\nu = x,y$ in 2D. 
By parity symmetry (spatial inversion), the off-diagonal quadrupole component strictly vanishes, i.e., $Q_{k_{x}k_{y}}=0$. 
Furthermore, under a $\pi/2$ spatial rotation $\bm{k}\rightarrow R\bm{k}$, the Brillouin zone of the square lattice remains invariant.
By utilizing the spectral symmetry dictated by the altermagnetic order, $\xi_{\sigma}(R\bm{k})=\xi_{-\sigma}(\bm{k})$ in Eq.~(\ref{symm_spectrum}), 
and performing a change of integration variables, one can show that the diagonal components satisfy:
\begin{equation}
Q_{k_{y}^{2}}= -Q_{k_x^{2}}. 
\label{relation_quad}
\end{equation}
This antisymmetric relation is the microscopic origin of the $d$-wave macroscopic response, directly mirroring the discrete symmetry found in the spin susceptibility Eqs.~(\ref{symm_chi},\ref{realtion_chi}).

To quantitatively establish the relationship between the linear response susceptibility and this intrinsic ground-state property, we evaluate $Q_{k_x^{2}}$ at zero temperature.
Integrating over the spin-split elliptical Fermi seas yields the analytical result:
\begin{equation}
Q_{k_x^{2}}  =\frac{k_{f}^{4}}{16\pi\sqrt{1-Q^{2}}}
\Big(\frac{1}{1+Q} -\frac{1}{1-Q} \Big)
\end{equation}
Comparing this intrinsic quadrupole moment with the expression for the cross-susceptibility derived in Eq.~(\ref{eq:chi}), we uncover an algebraic relation:
\begin{equation}
\chi = -\frac{16m}{G^{2} k_{f}^{2}} Q_{k_x^{2}}.
\end{equation}
The induced spin-charge cross-susceptibility is proportional to the intrinsic spin quadrupole moment in momentum space.
Both quantities increase monotonically with the $d$-wave interaction strength and vanish identically in the normal phase ($Q=0$).
From a physical standpoint, while $Q_{k_x^{2}}$ serves as the microscopic order parameter that breaks time-reversal and rotational symmetries, the cross-susceptibility $\chi$ acts as its macroscopic thermodynamic observable. 
Measuring this anisotropic spin response to a scalar charge potential provides a direct probe of the hidden $d$-wave altermagnetic order.

\textit{Discussion.}
We briefly discuss the universality and limits of the established relationship between the cross-susceptibility $\chi$ and the momentum-space spin quadrupole $Q_{k_x^2}$. 
As demonstrated in our preceding analysis, the linear proportionality emerges asymptotically in the low-electron-density limit ($k_f \ll G$). 
If we phenomenologically generalize the scattering wavevector from the principal reciprocal lattice vectors to an arbitrary direction $\bm{q} = (q_x, q_y)$, symmetry considerations show that the spin response 
$\chi(\bm {q})$ satisfies 
$\chi(q_x, q_y)=-\chi (\pm q_y, q_x)$.
At the leading order, $\chi(\bm {q})$ exhibits a $d$-wave tensorial form factor, qualitatively proportional to $q_x^2 - q_y^2$. 
This provides an intuitive explanation for the spatial anisotropy of the induced spin density: A periodic potential modulated along the diagonal direction ($q_x = \pm q_y$) would decouple from the $d$-wave spin nematicity, inducing zero net spin response. 

Furthermore, as the system moves away from the strictly dilute limit (i.e., for larger $k_f$), the higher-order terms in the Taylor expansion of the energy denominator can no longer be safely truncated. 
These $\mathcal{O}((k_f/G)^4)$ corrections naturally introduce secondary couplings to higher-order momentum-space spin multipoles (such as the spin hexadecapole). 
While the strict linear algebraic identity is a pristine feature of the truncated low-density regime, the robust qualitative correlation remains universally valid. 
Even in the presence of higher-order multipole contributions, the spin-charge cross-response is driven by the strength of the intrinsic $d$-wave order parameter, reflecting the underlying spin quadrupole moment of the altermagnetic ground state.

\textit{Conclusion.}
In summary, we have theoretically investigated the spin-charge cross-response in a 2D $d$-wave UM (altermagnetic) state subjected to a weak periodic lattice potential. 
By employing the Matsubara Green's function formalism within the low-electron-density regime, we analytically derived the static cross-susceptibility. 
Our results demonstrate that a purely non-magnetic scalar potential can induce a spatially periodic real space spin 
distribution.
In addition, we have revealed that the induced susceptibility is proportional to the intrinsic momentum-space spin quadrupole moment. 
Furthermore, symmetry analyses indicate that this cross-response acts as a tensorial, directional probe, offering a direct signature of the hidden $d$-wave spin-splitting.

While our continuum model with an empty-lattice approximation successfully captures the 
universal symmetry properties, extending this paradigm to tight-binding models exhibiting 
$d$-wave UM is highly desirable. 
Such an extension, particularly by incorporating multi-orbital degrees of freedom, would bridge the gap between our phenomenological framework and relastic systems. 
Crucially, it would enable direct comparisons with emerging experimental observations, where the intricate interplay between unconventional magnetic orders and anisotropic Fermi surface reconstructions plays an important role.

\textit{Acknowledgments.} 
We are grateful to the stimulating discussions with Zhuang Qian. 
C.W. is supported by the National Natural Science Foundation of China under the Grants No. 12234016 and No. 12174317.  
Z.P. is supported by the National Natural Science Foundation of China under Grant No. 12504219 and the startup funding from Xiamen University. 
This work has been supported by the New Cornerstone Science Foundation.


\appendix

\setcounter{equation}{0}
\setcounter{figure}{0}
\setcounter{table}{0}
\renewcommand{\thefigure}{A\arabic{figure}}

\renewcommand*{\theHfigure}{\thefigure}

\section{Detailed Derivative}
In this appendix, we provide the detailed derivation of the static spin-charge cross-susceptibility $\chi(\bm{G}_1)$ using the Matsubara formalism. 
We start from the continuum low-energy effective Hamiltonian for the $d$-wave unconventional magnetic (altermagnetic) state, subjected to a weak periodic lattice potential.
In real space, the unperturbed Hamiltonian and the scalar crystal potential are given respectively by:
\begin{align*}
&H_{0} =  -\frac{1}{2m}[\nabla^{2} +Q(\partial_{x}^{2}-\partial_{y}^{2})\sigma_{z}]-\mu,    \\
&V(x,y) =  V_{0} (\cos Gx+\cos Gy),
\end{align*}
where $\mu=k_{f}^{2}/2m$ is the chemical potential characterized by the effective Fermi momentum $k_f$, and $Q$ is the dimensionless 
$d$-wave order parameter.
Upon Fourier transforming to momentum space, the unperturbed Hamiltonian becomes block-diagonal in the spin basis:
\begin{align*}
H_{0}(\bm{k}) & = \frac{1}{2m}[\bm{k}^{2}+Q(k_{x}^{2}-k_{y}^{2})\sigma_z-k_{f}^{2}],
\end{align*}
where the Pauli matrix $\sigma_z$ has eigenvalues $\sigma=\pm 1$ for spin-up and spin-down electrons. 
The periodic potential transfers momentum precisely by the primitive reciprocal lattice vectors $\bm{G}_1=(G,0)$ and $\bm{G}_2=(0,G)$:
\begin{align*}
V(\bm{q}) & = \frac{V_{0}}{2}\sum_{\bm{G} \in \{\pm\bm{G}_1, \pm\bm{G}2\}}\delta_{\bm{q},\bm{G}}.
\end{align*}
The non-interacting Matsubara Green's function for the system is a diagonal $2\times 2$ matrix in spin space, $G_{0}(\bm{k},i\omega_{n})=  [i\omega_{n}-H_{0}]^{-1}$, which can be written as:
\begin{align*}
G_{0}(\bm{k},i\omega_{n}) & =\frac{1}{2} (
\frac{1+\sigma_{z}}{i\omega_{n}-\xi_{\uparrow}(\bm{k})}
+\frac{1-\sigma_{z}}{i\omega_{n}-\xi_{\downarrow}(\bm{k})} ),
\end{align*}
where the spin-split energy dispersions are:
\begin{equation*}
\begin{aligned}
\xi_{\uparrow}(\bm{k}) & = \frac{1}{2m}(\frac{k_{x}^{2}}{a^{2}}+\frac{k_{y}^{2}}{b^{2}}-k_{f}^{2}),\\
\xi_{\downarrow}(\bm{k}) & = \frac{1}{2m}(\frac{k_{x}^{2}}{b^{2}}+\frac{k_{y}^{2}}{a^{2}}-k_{f}^{2}).
\end{aligned}
\end{equation*}
Here, we have introduced the dimensionless scaling parameters, $a^{-2}=1+Q$ and $b^{-2}=1-Q$ to demonstrate the elliptical geometry of the spin-dependent Fermi surfaces, from which the order parameter can be parameterized as $Q=\frac{1}{2}(a^{-2}-b^{-2})$.

To evaluate the induced spin response, we calculate the static cross-susceptibility bubble $\chi(\bm{G}_1)$ shown in Fig.~\ref{fig:Bubble},
\begin{equation*}
\begin{aligned}
\chi(\bm{G}_{1})  = T\sum_{i\omega_{n}}\int\frac{d ^{2}\bm{k}}{(2\pi)^{2}} \text{tr}[G(\bm{k},i\omega_{n})G(\bm{k}+\bm{G}_{1},i\omega_{n})\sigma_{z}].
\end{aligned}
\end{equation*}
Tracing over the spin degrees of freedom and take the Matsubara summation, we obtain
\begin{equation*}
\begin{aligned}
\chi(\bm{G}_{1})& =  \int_{\text{BZ}}\frac{d ^{2}\bm{k}}{(2\pi)^{2}} \sum_{\sigma} 
\sigma\frac{n_{F}[\xi_{\sigma}(\bm{k})]-n_{F}[\xi_{\sigma}(\bm{k}+\bm{G}_{1})]}{\xi_{\sigma}(\bm{k})-\xi_{\sigma}(\bm{k}+\bm{G}_{1})}.
\end{aligned}
\end{equation*}
This final form provides the analytical basis for evaluating the spin-charge cross-response at any temperature.

At zero temperature, the Fermi-Dirac distribution reduces to a step function, $n_{F}(\xi_{\sigma\bm{k}})=\Theta(-\xi_{\sigma\bm{k}})$.
In the relevant metallic regime with low electron density, the Fermi momentum is small, $|\bm{k}|\ll G/2$.
The integration completely avoids the singularities at the Brillouin zone boundaries.
By multiplying the numerator and denominator by appropriate factors, we can decompose the cross-susceptibility into four constituent integrals representing the particle and hole contributions from both spins:
\begin{equation}
\begin{aligned}
\chi(\bm{G}_{1}) 
=&  2m(-a^{2}I_{1}+a^{2}I_{2}+b^{2}I_{3}-b^{2}I_{4}),
\end{aligned}
\end{equation}
where the integrals are explicitly given by,
\begin{equation}
\begin{aligned}
I_{1}(a,b) & = \int_{\text{BZ}}\frac{d ^{2}\bm{k}}{(2\pi)^{2}}\frac{\Theta[-\xi_{\uparrow}(\bm{k})]}{2\bm{k}\cdot\bm{G}_{1}+\bm{G}_{1}^{2}},   \\
I_{2}(a,b) & = \int_{\text{BZ}}\frac{d ^{2}\bm{k}}{(2\pi)^{2}}\frac{\Theta[-\xi_{\uparrow}(\bm{k}+\bm{G}_{1})]}{2\bm{k}\cdot\bm{G}_{1}+\bm{G}_{1}^{2}},    \\
I_{3}(a,b) & = \int_{\text{BZ}}\frac{d ^{2}\bm{k}}{(2\pi)^{2}}\frac{\Theta[-\xi_{\downarrow}(\bm{k})]}{2\bm{k}\cdot\bm{G}_{1}+\bm{G}_{1}^{2}},  \\
I_{4}(a,b) & = \int_{\text{BZ}}\frac{d ^{2}\bm{k}}{(2\pi)^{2}}\frac{\Theta[-\xi_{\downarrow}(\bm{k}+\bm{G}_{1})]}{2\bm{k}\cdot\bm{G}_{1}+\bm{G}_{1}^{2}}.
\end{aligned}
\end{equation}
where the parameterizations are defined as $a^{-2}=1+Q$ and $b^{-2}=1-Q$.

To calculate the unshifted integral $I_{1}$, we introduce the scale transformation $k_{x}=aq\cos\theta$ and $k_{y}=bq\sin\theta$, such that the energy dispersion becomes isotropic in the scaled coordinates: $\xi_{\uparrow}=\frac{1}{2m}(q^2-k_f^2)$.
The integration measure is $d^2\bm{k}=ab q dq d\theta$. 
This yields:
\begin{equation*}
\begin{aligned}
I_{1}(a,b) & = \frac{ab}{2\pi}\int_{0}^{k_{f}}qd q\int_{0}^{2\pi}\frac{d \theta}{2\pi}\frac{1}{2aq\cos\theta G+G^{2}}\\
& = \frac{ab}{2\pi}\int_{0}^{k_{f}}\frac{q}{G\sqrt{G^{2}-4a^{2}q^{2}}}d q\\
& = \frac{b}{8\pi a}[1-\sqrt{1-a^{2}(\frac{k_{f}}{G/2})^{2}}],
\end{aligned}
\end{equation*}
For the shifted integral $I_{2}$, which represents the hole contribution, the parameterization shifts the origin: $k_x=aq\cos\theta-G$ and $k_y=bq\sin\theta$.
Evaluating the angular integral introduces a negative sign due to the geometry of the shifted denominator, rigorously yielding $I_2(a,b)=-I_1(a,b)$.

By symmetry, the spin-down integrals are obtained simply by exchanging the parameters, $I_3(a,b)=I_1(b,a)$ and $I_4(a,b)=-I_1(b,a)$.
Summing these contributions provides an exact analytical expression for the cross-susceptibility valid for arbitrary elliptical Fermi surfaces (as long as they do not touch the zone boundary):
\begin{equation*}
\begin{aligned}
& \chi(\bm{G}_{1}) = 2m(-a^{2}I_{1}+a^{2}I_{2}+b^{2}I_{3}-b^{2}I_{4}) \\
& =  \frac{mab}{2\pi}\Bigg\{\sqrt{1-a^{2}(\frac{k_{f}}{G/2})^{2}}-\sqrt{1-b^{2}(\frac{k_{f}}{G/2})^{2}}\Bigg\}.
\end{aligned}
\end{equation*}

To elucidate the role of the $d$-wave order, we substitute $a$ and $b$ with the order parameter $Q$. 
Using $a^{-2}=1+Q$ and $b^{-2}=1-Q$, the result can be rewritten and Taylor-expanded for a weak $d$-wave UM phase:
\begin{equation*}
\chi(\bm{G}_{1}) 
= \frac{m}{2\pi}\frac{(2k_{f}/G)^{2}}{\sqrt{1-(2k_{f}/G)^{2}}}Q+\mathcal{O}(Q^{2}).
\end{equation*}
This demonstrates that the spin response is fundamentally driven by and is linearly proportional to the $d$-wave order $Q$.

In the low-density limit $k_f\ll G/2$, a Taylor expansion with respect to $(k_f/G)$ gives the leading order behavior:
\begin{equation*}
\begin{aligned}
\chi(\bm{G}_{1}) 
\approx -\frac{m}{4\pi} \Big(\frac{k_{f}}{G/2}\Big)^{2} ab(a^{2}-b^{2}).
\end{aligned}
\end{equation*}
Remarkably, this result encodes a profound physical quantity. 
The intrinsic momentum-space spin quadrupole moment of the altermagnetic Fermi sea is defined as:
\begin{equation*}
\begin{aligned}
Q_{k_x^{2}}\equiv& \int_{\text{BZ}}\frac{d ^{2}\bm{k}}{(2\pi)^{2}} k_x^2 
\{ n_{F}[\xi_{\uparrow}(\bm{k})]-n_{F}[\xi_{\downarrow}(\bm{k})]\}.
\end{aligned}
\end{equation*}
By performing the integration over the respective elliptical areas, we exactly find 
\begin{equation*}
\begin{aligned}
Q_{k_x^{2}}=& \frac{k_{f}^{4}}{16\pi}ab(a^{2}-b^{2}).
\end{aligned}
\end{equation*}
Thus, the induced cross-susceptibility is directly proportional to the spin quadrupole moment of the unperturbed state:
\begin{equation*}
\chi(\bm{G}_{1}) = -\frac{16m}{G^{2}k_f^2} Q_{k_x^{2}}.
\end{equation*}

Finally, as proven in the main text, the symmetry of the spectrum $\xi_{\sigma}(R\bm{k})=\xi_{-\sigma}(\bm{k})$ under a $\pi/2$ rotation guarantees that $\chi(R\bm{G}_1)=-\chi(\bm{G}_1)$.
Utilizing $\bm{G}_2=R\bm{G}_1$, the first-order correction to the real-space spin density is assembled as:
\begin{align*}
S_z(\bm{r}) =&\sum_{\bm{q}\in\{\pm \bm{G}_1,\pm\bm{G}_2\}} 
\chi(\bm{q}) \frac{V_0}{2} e^{i\bm{q}\cdot\bm{r}}   \\
=&\chi(\bm{G}_1) V_0 \big( \cos Gx -\cos Gy\big).
\end{align*} 
This links the microscopic $d$-wave geometry of the Fermi surface to the macroscopic spatial modulation.

\twocolumngrid

\bibliography{references}

\end{document}